# Design of a Power Amplifying-RIS for Free-Space Optical Communication Systems

Alain R. Ndjiongue, Telex M. N. Ngatched, Octavia A. Dobre, and Harald Haas

*Abstract*—The steering dynamics of re-configurable intelligent surfaces (RIS) have hoisted them to the front row of technologies that can be exploited to solve skip-zones in wireless communication systems. They can enable a programmable wireless environment, turning it into a partially deterministic space that plays an active role in determining how wireless signals propagate. However, RIS-based communication systems' practical implementation may face challenges such as noise generated by the RIS structure. Besides, the transmitted signal may face a double-fading effect over the two portions of the channel. This article tackles this double-fading problem in near-terrestrial free-space optical (nT-FSO) communication systems using a RIS module based upon liquid-crystal (LC) on silicon (LCoS). A doped LC layer can directly amplify a light when placed in an external field. Leveraging on this capacity of a doped LC, we mitigate the double-attenuation faced by the transmitted signal. We first revisit the nT-FSO power loss scenario, then discuss the direct-light amplification, and consider the system performance. Results show that at $51^o$ of the incoming light incidence angle, the proposed LCoS design has minimal RIS depth, implying less LC's material. The performance results show that the number of bit per unit bandwidth is upper-bounded and grows with the ratio of the sub-links distances. Finally, we present and discuss open issues to enable new research opportunities towards the use of RIS and amplifying-RIS in nT-FSO systems.

## I. INTRODUCTION

With the on-going deployment of 5G and the upcoming 6G, full coverage of the environment is one of the essential targets. Among the solutions and techniques used to mitigate skip-zones in wireless communication systems, re-configurable intelligent surfaces (RIS) emerge as excellent keys to solve the dead-zone dilemma. A RIS module can be used to increase the signal-to-noise ratio at a specific user, especially when the line-of-sight signal is not guaranteed. Consequently, the data rate and communication range can be improved for the desired user. The RIS technology also enhances the communication channel by making it smart and improving the communication network capacity. It is also coined that a RIS infrastructure increases the spectrum and energy efficiencies [1], [2]. RIS modules can be exploited in systems with a variety of applications in radio frequency and optical communications [3], [4]. Practically, RISs are two-dimensional, nearly passive or active surfaces of material that are electronically controllable with integrated electronics and possess a small thickness. Controlling the electromagnetic and physical characteristics of a RIS structure enables the module to operate in two different scenarios depending on the system architecture. Especially, the RIS module can be exploited to reflect and redirect the incoming signal towards the desired receiver or transmit the incident wave to form a refractive system. In both cases, the RIS module steers the incoming signal. These advantages have provided the RIS technology with a significant role in solving the skip-zone's problem in wireless communication systems [5], [6] and improving the field-of-view in visible light communication receivers [7].

Based on the above brief introduction, RIS-based wireless communication systems are clearly beneficial. However, their implementations faces a couple of challenges. If we consider reflective RIS modules, up to 10% of the incoming signal power may be lost during the transition through the RIS structure [7]. Moreover, the RIS structure may generate noise. Furthermore, the transmitted wave faces a double-fading situation due to the sub-channels before and after the module. These attenuation phenomena may significantly affect the transmitted signal. A potential solution to this dilemma is to design a RIS module with amplifying capabilities. In near-terrestrial free-space optical (nT-FSO) communication systems, a power amplifying-RIS module can also be designed to solve the double-fading effects generated by using the RIS technology. The latter represents the focus of this article.

Because the communication signal in nT-FSO is encrusted in a light, the RIS material used in such environments must have the potential to reflect or refract lights. From this perspective, in refraction, the material should have excellent refractive capabilities with minimized reflectiveness. Similarly, to minimize internal losses in reflection, the material should be a perfect mirror for the incoming light with a low transmittance. Such mirrors have a reflective coefficient close to unity and a null refractive coefficient. The RIS technology has been introduced to FSO systems in [8], where a statistical channel model for the geometric and misalignment losses was proposed. It is clear that the transmitted signal faces the turbulence, pointing effects, and losses twice. Lately, doped liquid-crystals (LCs) have been presented as an excellent material for VLC receivers to steer the incoming light, improve the receiver field-of-view, and enhance signal reception [7]. It was also demonstrated that a doped LC module could simultaneously steer and amplify the incoming light for better reception, and improve the transmission range. This article exploits the LCs' refractive and amplification properties and the CMOS silicon backplane reflection attributes to design a power

A. R. Ndjiongue, T. M. N. Ngatched, and O. A. Dobre are with the Faculty of Engineering and Applied Science, Memorial University of Newfoundland, Canada.
Harald Haas is with the LiFi Research and Development Center, Dpt. Electronic and Electrical Engineering, the University of Strathclyde, Glasgow, U.K.

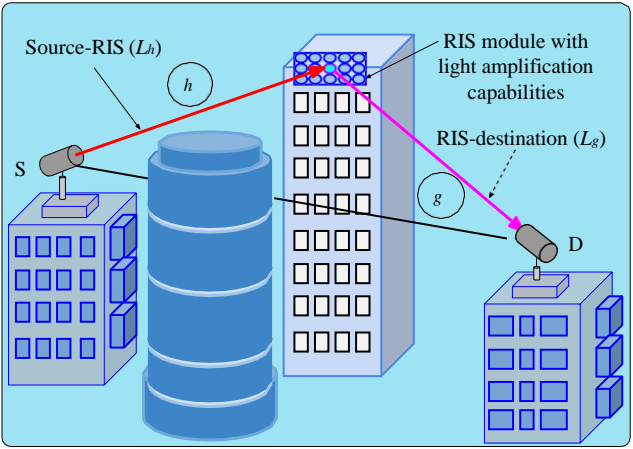

Fig. 1: Model of nT-FSO link using a RIS module with power amplification capabilities.

amplifying-RIS module for nT-FSO systems. The obtained combined structure, LCoS, embodies four main attributes, including refraction, amplification, reflection, and chromatic correction [9]. The use of LCoS-based RIS in nT-FSO presents a threefold advantage: ($i$) enhancing the signal strength at the receiver; ($ii$) reducing noise propagation, and ($iii$) solving chromatic dispersion in multi-color transmissions.

Initially, the LCoS technology was developed for image and video display applications by exploiting the light-modulating attributes of LC materials and CMOS silicon backplane's advantages to realize a transmissive device. The obtained structure impacts the incident light's phase and amplitude in both reflection and refraction phenomena by tuning the LC properties and parameters, including the birefringence, temperature, and refractive index using an externally applied electric or magnetic field. In this article, we combine a doped LC and CMOS silicon backplane to design a power amplifying-RIS, which is exploited to reflect the incident light. We summarize our contributions as follows: ($i$) we design an LCoS-based power amplifying-RIS for nT-FSO systems; ($ii$) we derive optimal RIS dimensions, reducing the LC material required, and thus lowering the cost; ($iii$) based on the nT-SFO link equation, we analyze both spectral and energy efficiencies of the system, and provide performance results; ($iv$) finally, we discuss open issues and implementation challenges related to the proposed design, leading to new research directions.

## II. POWER LOSS SCENARIOS IN nT-FSO

In this section, we discuss sources of power loss in nT-FSO communication systems.

We consider the nT-FSO system illustrated in Fig. 1, where it is required to send a message from one building source, S, to a destination building, D. Because the optical signal in space is a directed point-to-point light, the best link is the line-of-sight signal from S to D. However, this direct S-D link is obstructed by an obstacle. An RIS module is deployed on a surrounding building's facade to assist the source in overcoming the disadvantageous propagation conditions by providing a high-quality virtual link from S to D.

### A. RIS-based nT-FSO Systems

Unlike wireless communication free-space channels using the radio frequency, nT-FSO channels do not have a non-line-of-sight path. As a result, as depicted in Fig. 1, with the line-of-sight signal cut-off, the receiver is totally out of reach. The virtual line-of-sight link enabled by the RIS has two sub-links. These sub-links are likely to be affected by most of the impairment sources presented in the sequel. Among these, most loss coefficients impact the system twice. As an example, the distance-related losses affect both the S-RIS and RIS-D links. The propagation loss will be doubled depending on the S-RIS and RIS-D distances compared to the S-D distance and channel behavior over the two sub-links. In addition to the RIS reflection coefficient, which is in general less than unity, new equipment losses are created by introducing the RIS module, including geometric and optical losses. The proposed power amplifying-RIS aims at compensating for these losses.

### B. Defining nT-FSO Power Losses and Attenuation

Path loss affects the amplitude of the transmitted light when it travels through space. They may result from channel instability, the equipment, or techniques used in the system. They all induce a power loss, and consequently, reduce the system's energy efficiency. Generally, the received power is the channel transmittance product by the transmit power. This transmittance is due to the sum of losses, including propagation losses and the effect of turbulence-induced fading [10]. We discuss these losses in the following subsections.

*1) Path Loss:*
**Propagation losses:** These are losses related to light propagation through space, and include absorption and scattering losses. They sum up the attenuation coefficient as defined by the Beer-Lambert laws. However, these losses can be individually analyzed to evaluate each source's contribution as proposed in [11]. The absorption loss stems from light photons' interaction with molecules in the space through which the light propagates. This loss varies with the wavelength and the molecule concentration in the space.

On the other hand, the scattering loss is also generated from collision between light photons and the particles in space and can be classified as Rayleigh, Mie, and non-selective scatterings [11]. Their effects are intensified by rains, fogs, and clouds [12]. As defined by the Beer-Lambert laws, absorption and scattering are given in dB, while the attenuation coefficient is given in dB/m and represents the sum of absorption and scattering coefficients. Propagation loss can sometimes be combined with geometrical loss, as given in [10, Eq. (2)].
**Geometrical loss:** This loss defines the part of light falling outside the photodetector's (PD) surface because the generated light forms a Gaussian beam, which at the receiver has a diameter greater than the PD's diameter. It is expressed in terms of the PD's diameter, beam divergence at the transmitter, and transmission distance. Since the PD's diameter and beam divergence are constant, geometrical loss mostly varies with the transmission distance.
**Scintillation Loss:** In nT-FSO, turbulence in the transmission environment impedes the propagating signal, creating signal loss due to scintillation. This loss is generally expressed in terms of the structure's refractive index, wavelength, and transmission distance [11].

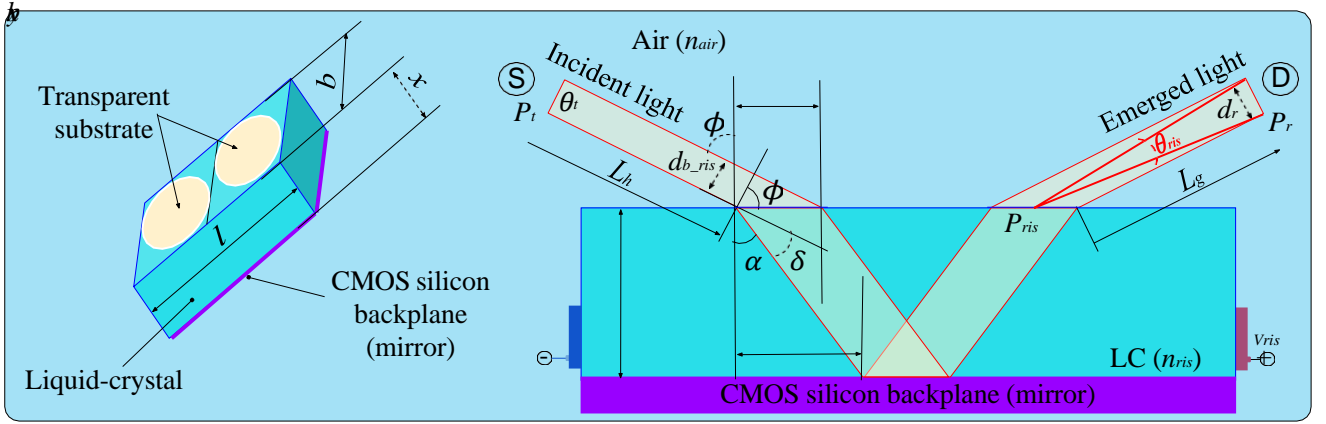

Fig. 2: Amplification principle and geometry of the proposed LCoS-based RIS. $P_t$ = transmit power, $\theta_t$ = beam divergence at S. $L_h$ = S-RIS distance, $\varphi$ = incidence angle, $\alpha$ = angle of refraction, $\delta$ = retardation angle, $x$ = RIS depth, $d_{b\_ris}$ = beam diameter at RIS, $P_{ris}$ = RIS emerged light power, $\theta_{ris}$ = virtual divergence at RIS, $L_g$ = RIS-D distance, $d_r$ = beam diameter at D, $b$ = spot diameter of glass substrate, and $l$ = RIS length. Note that $\varphi$, $\alpha$, and $\delta$ are measured on the incoming central ray.

**Pointing and optical losses:** A miss-alignment between the transmitting light source and PD leads to pointing losses. On the other hand, the optical equipment in the transmitter and receiver may create optical losses, which are expressed in terms of the transmitter and receiver's efficiencies.

*2) Attenuation:*
Notwithstanding the attenuation discussed above, rain, fog, and cloud also attenuate the transmitted light and constitute scattering sources. In general, rain is characterized by the rainfall rate and distance along which the signal is affected. These parameters help to define the rain attenuation. Fogs and clouds are generally included in the scattering coefficient or defined in terms of visibility.

## III. THE PROPOSED DESIGN

In this section, we describe the proposed RIS module, discuss its geometry and internal structure.

### A. Module Description

In [7], the authors proposed a direct-light amplifying module using doped LCs. This article exploits this doped LC's amplifying attributes to design a direct-light amplifying-RIS to solve the double-fading effect in nT-FSO systems. Figure 2 depicts the proposed RIS module (LCoS). On the left-hand side, the figure shows the perspective view of the module. The light penetrates the element through the glass substrate, and the amplified light leaves the RIS structure through the same glass substrate. On the right-hand side, the figure shows a profile-view highlighting the RIS geometry, giving an insight into the light path through the element. The RIS cell contains an LC material with nanodisk infiltration or any other light amplifying doping substance.

An LCoS has the layered LC architecture depicted in Fig. 3. Its structure is similar to that of conventional LCs except that the top layer is a thin glass substrate, while the bottom layer is a CMOS silicon backplane. The incident light travels through the LCoS structure with a reduced scattering effect and is amplified. While the externally applied field controls the LC properties, a high-performance circuitry deals with phase retardation of the wavefront. Hence, the LCoS-based RIS modulates both the amplitude and the phase of the incoming light. Its amplitude depends on both the linear polarizer and the externally applied field, while the phase shift is controlled by the adjustment of the refractive index.

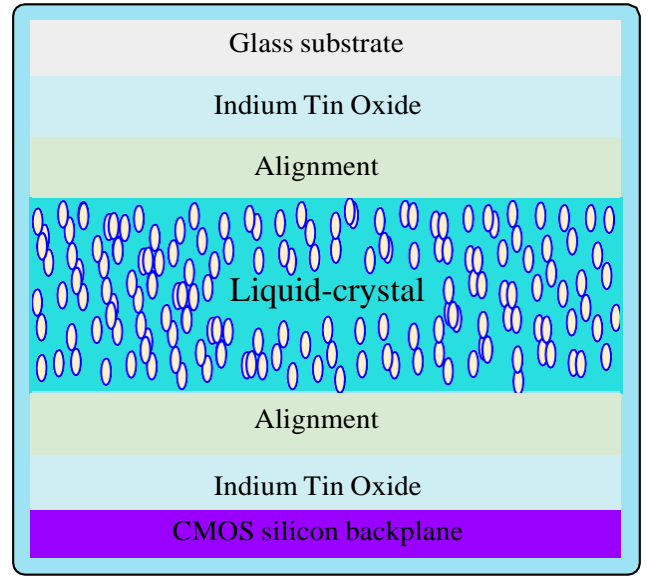

Fig. 3: Structure of the proposed LCoS-based RIS. It consists of a glass substrate, two Indium Tin Oxide layers, two alignment layers, the LC material, and a CMOS silicon backplane mirror.

### B. RIS Geometry

As shown in Fig. 2, the light beam before the RIS module is characterized by its transmit power, the divergence angle, the S-RIS distance, and the beam diameter at RIS. The light falls on the glass substrate with an incidence angle and is spread on the substrate surface (on the glass substrate) over a diameter, $b$. After the RIS module, the light travels over the RIS-D distance to reach the receiver. The light power emerging from the RIS structure is evaluated by the product of the RIS's transmittance and light power entering the RIS's structure.

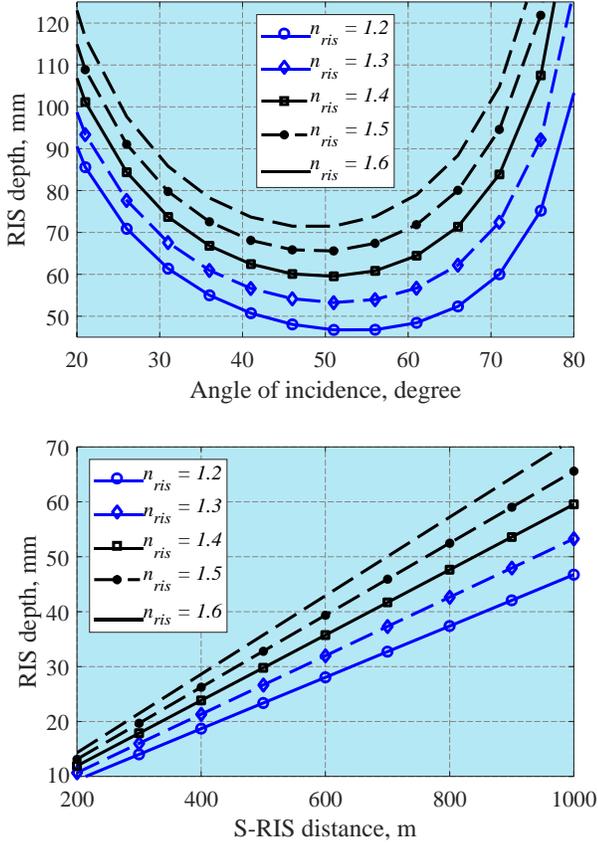

Fig. 4: RIS depth in terms of incidence angle (top); the S-RIS link distance for multiple values of the RIS refractive index (bottom).

**RIS depth optimization:** The incoming light, which faces refraction on the glass substrate, propagates through the RIS structure with a refraction angle due to the shift given by the retardation angle. The furthest-to-the-left ray reflects on the CMOS silicon backplane and may interfere with the rest of the incident beam outside or inside the RIS structure, depending on the RIS depth and length. To minimize this interference, we adopt the configuration described in Fig. 2, where the interference area between this ray and the incoming beam is minimized. It can be shown that the value of the RIS depth that minimizes this interference is lower-bounded as $x \geq \theta_t L_h / 2 \tan(a) \cos(\varphi)$. Because the device's tunability is mostly based on controlling the RIS electro-optic parameters such as the refractive index, the RIS depth is determined using this lower bound expression.

We consider a receiver made of an avalanche PD, with 2.5 mm diameter, which can handle about 1 GHz of bandwidth. We also consider a power amplifying-RIS module with the smallest depth. As shown in the bottom-part of Fig. 4, the calculated RIS depth, $x$, given in terms of the S-RIS distance, increases proportionally with both the S-RIS link distance and the RIS refractive index. In the top-part of Fig. 4, we show the same RIS depth in terms of the incidence angle for the same refraction performance. It can be observed that for an S-RIS link distance of 1000 m, a minimum RIS depth of 46.74 mm is obtained for an incidence angle of 51° and a RIS refractive index of 1.2. The value of the RIS depth increases with the refractive index, and reaches 71.49 mm when the refractive index is 1.6 for the same incidence angle of 51°. It is essential to mention that smaller values - close to 0° - and larger values - close to 90° - of the incidence angle should be avoided as they lead to high values of the RIS depth. The amount of LC material required increases with the RIS depth.

**The module casing:** The overall cover is opaque to avoid outside lights influencing the light traveling within the RIS structure and is made of anti-reflective materials to reduce inner reflections. The minimum value of the RIS depth is discussed above. As presented in Fig. 2, the spot on the glass substrate is defined by a diameter $b = \theta_t L_h / \cos(\varphi)$. This diameter dictates the RIS module's width, which is also $b$ without considering the module cover thickness. Similarly, the RIS minimum length is $3b$.

### C. RIS Internal Materials and Structure

**The liquid-crystal:** Several chemical materials can be used as LC. However, they must present a good amplification coefficient and refractive index tunability. Various types of off-the-shelf LC phases are available, including nematic and smectic phases, to cite only these two. Nevertheless, one of the essential LC properties is the birefringence, which is positive in general and evaluated as the difference of refractive indices due to perpendicular and parallel polarization of the externally applied field compared to the director. The director is a base vector parallel to the optical axis of the crystals. The phase retardation, given in the right-hand-side of Fig. 2, is the difference between the incidence and refracted angles, and can be calculated using the birefringence, wavelength, and the RIS thickness. Note that this thickness represents the intrinsic depth of the LC layer shown in Fig. 3, and its maximum value can be evaluated based on the maximum phase delay. It is crucial to find a trade-off between the values of the RIS depth calculated using the refractive index and those obtained from geometrical evaluation using Fig. 2.

The RIS depth and crystal orientation can be adjusted so that the light reflected on the CMOS silicon backplane is oriented towards the receiver's direction. This simultaneously adjusts the RIS transmittance for a better signal-to-noise ratio, transmission rate, and service quality. Tuning can be performed using the externally applied voltage, governed by an efficient algorithm.

**CMOS silicon backplane:** The RIS aperture is a transparent glass substrate with a low reflection coefficient to reduce losses due to the transition from air to the RIS structure and vice-versa. On the other hand, the CMOS silicon backplane, which represents the bottom layer as shown in Fig. 3, is made of a material with high reflection coefficient. In the current applications of LCoS, the performance of traditional silicon backplanes is improved by the CMOS process. Its top-face must provide high reflectivity to reflect the incident light with less loss. Most available CMOS silicon backplanes are provided for wavelength from 180 nm to about 1550 nm and can be tuned with 1.8 V supply [13].

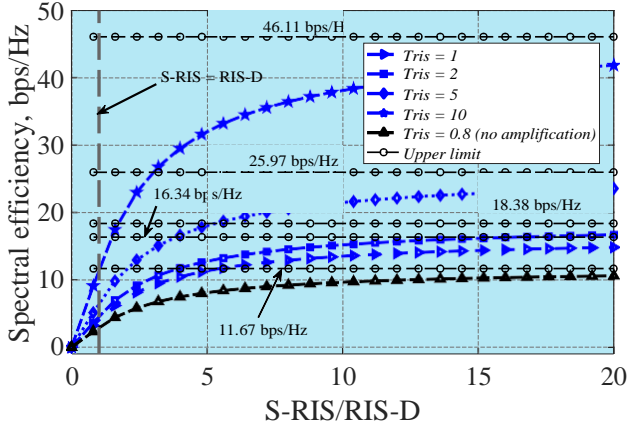
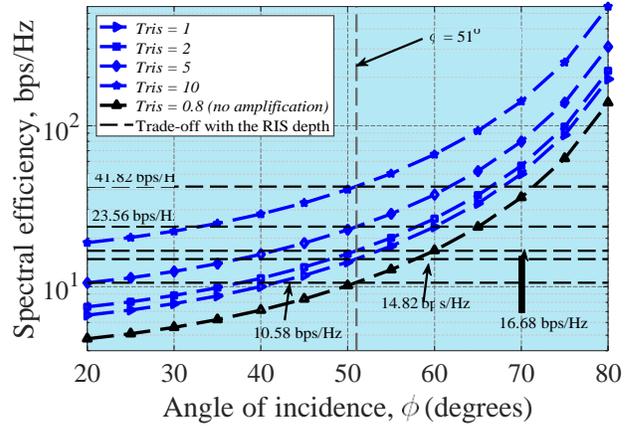

Fig. 5: Spectral efficiency of the system in terms of the distance ratio (left), and in terms of the incidence angle (right) for selected values of the RIS' transmittance.

## IV. ENERGY AND SPECTRAL EFFICIENCIES

This section discusses the overall system's energy and spectral efficiencies, and highlights the advantages of using a power amplifying-RIS in nT-FSO systems.

We show that the RIS position, the distances S-RIS and RIS-D, and their ratio are essential in the system performance. This is demonstrated through the spectral efficiency given as a function of the distance ratio. The results depicted in Fig. 5 consider the following scenario. The transmitter generates light with a wavelength of 780 nm, at a transmit power of 0.3 W, with an output efficiency of 95% and a beam divergence of 1 mrad. The generated light travels through an nT-FSO space over the S-RIS distance between 0 and 1 km before reaching the RIS module. The glass substrate is characterized by a 95% efficiency (loss in the reflection scenario of 5% of the power arriving at the glass substrate). Due to the RIS's small thickness, it is assumed that the beam diameter does not change. Hence, it is defined by its minimum size given by the aperture diameter, $b$, width, $b$, thickness, $x$, and length, $3b$. As it approaches the RIS's surface, the beam diameter is evaluated as $\theta_t L_h$ due to the assumption on small-angle sine. However, due to the beam incline, it falls on the glass substrate with a diameter, $b$. The light emerging from the RIS structure keeps its chromaticity, and it is assumed that no light dispersion happens within the RIS structure. The light power experiences another 5% loss as it crosses the glass substrate from the RIS structure to the air. The receiver is situated at the RIS-D distance from the RIS module, also defined between 0 and 1 km. A 95% efficiency also characterizes the receiver, and its diameter is 2.5 mm. Since the RIS structure does not change the light shape, the beam diameter at D can be evaluated using the transmitter divergence and the total transmission distance

(S-RIS + RIS-D). Simultaneously, the PD perceives the signal as if it was generated at the RIS with a divergence defined as shown in Fig. 2. The light undergoes a total attenuation of 2.6 dB per link, proportionally to the transmission distance.

### A. Spectral Efficiency

We calculate the rate at which information can be transmitted per unit bandwidth, using the model given in [11, Eq. (3)], which expresses the data transmission (in bps) over an nT-FSO vertical link. The derived transmission rate, expressed in terms of distance ratio, S-RIS distance, bandwidth, and incidence angle, is successively divided by the transmission bandwidth and total power to obtain the spectral and energy efficiencies, respectively.

Next, we discuss the spectral efficiency of the proposed amplifying-RIS-based nT-FSO system. We evaluate the system's spectral efficiency with and without amplification for several values of the RIS transmittance. Selected values of the RIS transmittance are 0.8, 1, 2, 5, and 10. In this evaluation, we consider the ratio of distances S-RIS to RIS-D. This ratio drastically impacts the transmission efficiency, as shown in Fig. 5, where we depict the spectral efficiency in terms of both the distance ratio and incident angle for the selected values of the RIS transmittance.

The left-part of Fig. 5, obtained for a $51°$ incidence angle, shows that the amount of information transmitted per unit bandwidth and expressed in terms of the distance ratio rapidly increases from 0 bps/Hz to a high value as the distance ratio increases from 0 to 1. After 1, the slope is preserved until the spectral efficiency's upper bound, regardless of whether the RIS is amplifying or not. In the non-amplifying-RIS case (RIS transmittance = 0.8), as the distance ratio approaches high values, the spectral efficiency approaches 11.67 bps/Hz. When amplification is applied, a similar system behavior is observed. But, we respectively obtain 16.34, 18.38, 25.97, and 46.11 bps/Hz of spectral efficiency for the selected values of the RIS transmittance (1, 2, 5, and 10). As the distance ratio increases, the spectral efficiency converges to its upper bound, which is smaller for lower values of the RIS transmittance.

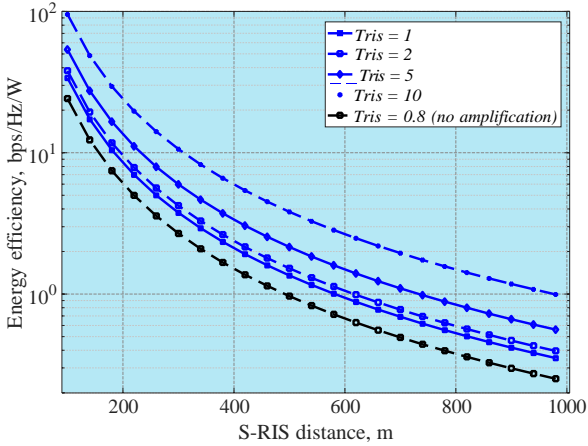

Fig. 6: The energy efficiency of the proposed power amplifying-RIS-based nT-FSO system in terms of the S-RIS distance, for selected values of the RIS transmittance (0.8, 1, 2, 5, and 10).

As an illustration, for a RIS transmittance of 0.8, a spectral efficiency upper bound of 11.67 bps/Hz is obtained at the distance ratio of 80. Whereas, for a RIS transmittance of 1, a spectral efficiency upper bound of 16.37 bps/Hz is obtained at the distance ratio of 100.

It is important to note that for all the proposed RIS transmittance values, the achieved spectral efficiencies are higher than those obtained from the non-amplifying-RIS. It should be noted that the distances S-RIS and RIS-D and their ratio are essential in a RIS-based nT-FSO system. This is valid irrespective of the angle of incidence, RIS transmittance, and other system parameters.

The right-part of Fig. 5 depicts the spectral efficiency as a function of the angle of incidence. It increases exponentially with the incidence angle. One would prefer to design a system with high values of the incidence angle for high range maneuverability. Nevertheless, this will also exponentially increase the size of the RIS module, especially its depth as shown in the top-part of Fig. 4, and will consequently lead to high cost. For this reason, we would prefer to keep the incidence angle around $51^o$, which is in the range of incident angle values offering smaller sizes of the RIS thickness (see Fig. 4). At these values of the incidence angle, we can verify that the obtained spectral efficiencies are below the upper limits given in the left-part of Fig. 5, which were obtained for a $51^o$ incidence angle. The spectral efficiency values are 10.58, 14.82, 16.68, 23.56, and 41.82 bps/Hz, which are respectively below 11.67, 16.34, 18.38, 25.97, and 46.11 bps/Hz. Conveniently, the incidence angle limitation keeps the bandwidth around 1 GHz to accommodate 5G and 6G systems, and the related transmission techniques, including multiple-inputs multiple-outputs, multiple access, and optical orthogonal frequency division multiplexing, to mention only these three.

### B. Energy Efficiency

Considering that energy represents one of the main factors required to transmit data, in Fig. 6 we show the impact of the S-RIS distance on the system's energy efficiency. It is seen that this decreases with the S-RIS distance of the S-RIS link. The system transmits fewer bits per unit of power consumption as the distance increases. For example, at 100 m S-RIS distance, the system can transmit up to 24.12 bps/Hz/W for a non-amplifying RIS. At the same distance, the power amplifying-RIS will transmit 33.77, 37.99, 53.67, and 95.27 bps/Hz/W for the selected RIS transmittance (1, 2, 5, and 10), respectively. These values decrease exponentially, and at 980 m S-RIS distance, we will have 0.25, 0.35, 0.39, 0.56, and 0.99 bps/Hz/W of energy efficiency, respectively, for a non-amplifying-RIS (transmittance = 0.8) and amplifying RISs with transmittance values of 1, 2, 5, and 10.

## V. OPEN CHALLENGES AND FUTURE RESEARCH

As with other wireless communication technologies, the implementation of RIS and amplifying-RIS-based nT-FSO systems raises a couple of issues. These challenges must be addressed to harness the full potential of this technology. To this end, we provide the main issues from our observation in the following paragraphs.

**Noise generation:** In direct-light amplification modules, as in the proposed LCoS-based RIS, the primary source of noise may be the spontaneous emission process, which has a spectrum similar to that of the gain amplifier. This noise is generated from the LC cavity fluctuation and pump source. Its intensity varies with the gain medium level. In doped amplifiers, the minimum noise figure can reach 8 dB in practical implementation. In general, the noise level in an optical amplifier depends on both the optical gain and the level of carrier inversion in the gain medium. Four techniques can be exploited to minimize this noise: (*i*) applying a stronger input signal to enable less amplification gain for an output signal power. This solution is not realistic because the light reaching the LCoS module is already attenuated; (*ii*) letting the amplifier operate at a high excitation level; (*iii*) using a single-mode amplification; and (*iv*) using a low-noise pump. The last three methods require further studies. It is also essential to analyze the RIS impact on the different types of noise found in VLC receivers since this may affect the system's performance.

**Interference generation:** Apart from its tunability, reduced thickness, and binary structure, the CMOS silicon backplane is characterized by a smooth surface and high reflection coefficient. Nevertheless, the light falling on its surface follows the Lambertian back-reflection pattern. Even though the specularly reflected light is mostly more substantial than the rest of the reflected rays, their effects on the transmitted message remain crucial and may contribute to the LC cavity losses, reducing the emerged light power. This aspect requires comprehensive studies.

**Liquid-crystal tuning time:** Tuning LCs takes a lapse of time in the order of milliseconds, which varies with the LC cavity size and can be measured using the optical switching method. It decreases when the dye used is dispersed into the LC substance and varies with the LC type and cell temperature. At $30^oC$ for example, a pure nematic LC tuning time is about 18 ms. When doped with a 0.5 wt% or one wt% dye, this LC tuning time can decrease to 13.5 or 11 ms, respectively, and reach 2 ms when the LC cell temperature reaches $100^oC$ [14]. This tuning time is enough to slow down the transmission speed, which should be constant or improved to meet the 5G or 6G requirements. A fast response LCs at high temperature should reduce the RIS tuning time [15]. The impact of the LC tuning time on the system performance and the design of fast tuning LCs deserve exhaustive investigations.

**Implementation issues:** The nT-FSO link can be affected by weather condition which may impact the transmitted signal, preventing the receiver to correctly decode the received data. To combat the weather influence, one may use a RIS module with high transmittance. This solution will be efficient only if the weather condition affects only the RIS-D link. If both links are under worse weather conditions, then an adaptive algorithm can be designed to adjust the transmit power and RIS transmittance depending on the link affected. This aspect should be thoroughly considered.

## VI. Conclusion

This article has elaborated on a power amplifying-RIS for nT-FSO systems to solve the double-fading coefficient that the transmitted signal undergoes. The proposed design is based on a direct-light amplification using LCoS. In addition to the efficient reflectiveness of the CMOS silicon backplane, the proposed design exploits the undeniable advantages provided by LCs in steering and amplifying light. RIS-based nT-FSO systems have been introduced, followed by the power loss scenarios in nT-FSO systems. A power amplifying module has been designed and proposed, and its geometry, depth, length, width, LC material, and the CMOS silicon backplane have been described. The system was evaluated through its spectral and energy efficiencies. It was shown that the minimal RIS thickness is obtained at an incidence angle of $51^o$ for a RIS refractive index of 1.2 over an S-RIS distance of 1000 m. The spectral efficiency was shown to be upper bounded when expressed in terms of the S-RIS and RIS-D distance ratio, and the upper bound depends on the RIS transmittance. Finally, open issues have been presented and discussed to enable new research opportunities towards the use of RIS and amplifying-RIS in nT-FSO systems.

**Alain R. Ndjiongue** [S'14, M'18, SM'20] received the M.Eng. and D. Eng degrees in electrical and electronic engineering from the University of Johannesburg, South Africa, in 2013 and 2017, respectively. He served as a senior lecturer at the same university from 2019 to 2020 and is currently a senior researcher at Memorial University of Newfoundland, Canada. His research interests are in digital communications including power line and optical communications. He is an active reviewer for several high impact IEEE ComSoc journals and conferences. He was a 2017 and 2018 exemplary reviewer for IEEE Communication Letters, Optical Society of America, and is a 2019 top 1% peer reviewer in the Essential Science Indicators Research.

**Telex. M. N. Ngatched** [M'05, SM'17] is an associate professor at Memorial University, Canada. His research interests include 5G enabling technologies, visible light and power-line communications, optical communications for OTN, and underwater communications. He is an associate editor of the IEEE Open Journal of the Communications Society and the Managing Editor of IEEE Communications Magazine. He was the publication chair of IEEE CWIT 2015, an associate editor for IEEE Communications Letters from 2015 to 2019, as well as Technical Program Committee member and session chair for many prominent IEEE conferences in his area of expertise.

**Octavia A. Dobre** [M'05, SM'07, F'20] is a professor and research chair at Memorial University, Canada. She was a visiting professor at Massachusetts Institute of Technology, as well as a Royal Society and a Fulbright Scholar. Her research interests include technologies for 5G and beyond, as well as optical and underwater communications. She has published over 300 referred papers in these areas. She serves as the Editor-in-Chief (EiC) of the IEEE Open Journal of the Communications Society. She was the EiC of the IEEE Communications Letters, a senior editor and an editor with prestigious journals, as well as General Chair and Technical Co-Chair of flagship conferences in her area of expertise. She was a Distinguished Lecturer of the IEEE Communications Society and a fellow of the Engineering Institute of Canada.

**Harald Haas** [S'98, AM'00, M'03, SM'16, F'17] received the Ph.D. degree from the University of Edinburgh in 2001. He currently holds the Chair of Mobile Communications at the University of Strathclyde, and is the initiator, co-founder and Chief Scientific Officer of pureLiFi Ltd as well as the Director of the LiFi Research and Development Center at the University of Strathclyde. His main research interests are in optical wireless communications, hybrid optical wireless and RF communications, spatial modulation, and interference coordination in wireless networks.